%% file: paper.tex
\documentclass[epj,referee]{svjour}
\usepackage{epsfig}

\newcommand{\lam}{\mbox{$\rm \Lambda ^0$}}
\newcommand{\alam}{\mbox{$\rm \bar \Lambda ^0$}} 

\newcommand{\up}{^\uparrow}

\newcommand{\upu}{u^\uparrow}
\newcommand{\downu}{u^\downarrow}

\newcommand{\upq}{q^\uparrow}
\newcommand{\downq}{q^\downarrow}

\newcommand{\upd}{d^\uparrow}
\newcommand{\downd}{d^\downarrow}

\newcommand{\ups}{s^\uparrow}
\newcommand{\downs}{s^\downarrow}

\begin{document}
\date{\begin{flushright} hep-ph/0204206, CERN-TH/2002-080 \end{flushright}}
\title{Intrinsic Polarized Strangeness and $\lam$ Polarization in 
Deep-Inelastic Production}
\author{John~Ellis\inst{1} \and Aram~Kotzinian\inst{2,3} \and 
  Dmitry V.~Naumov\inst{3}}
\institute{CERN, Geneva, Switzerland \and Yerevan Physics Institute, 375036, Yerevan, Armenia \and JINR, Dubna, 141980, Russia}
\abstract{
We propose a model for the longitudinal polarization of $\lam$ baryons
produced in deep-inelastic lepton scattering at any $x_F$, based on
static $SU(6)$ quark-diquark wave functions and polarized intrinsic
strangeness in the nucleon associated with individual valence quarks. Free
parameters of the model are fixed by fitting NOMAD data on the
longitudinal polarization of $\lam$ hyperons in neutrino collisions. Our
model correctly reproduces the observed dependences of $\lam$ polarization
on the kinematic variables. Within the context of our model, the NOMAD
data imply that the intrinsic strangeness associated with a valence
quark has anticorrelated polarization. We also compare our model
predictions with results from the HERMES and E665 experiments using 
charged leptons. Predictions of our model for the COMPASS experiment are 
also presented.
}
\maketitle
\input{intro}

\input{method}

\input{results}

\input{conclusion}

\begin{acknowledgement}

We are grateful to G. Ingelman and T. Sj\"ostrand for useful discussions
of the {\tt LEPTO} and {\tt JETSET} event generators.  We thank A. V.  
Efremov, M. G.  Sapozhnikov, R. Jaffe, Zuo-tang Liang and O. Teryaev for
very interesting and valuable discussions. D. N. is grateful to all
members of the NOMAD Collaboration, with special thanks to his Dubna
collaborators in the $\lam$ polarization analysis: S. A. Bunyatov, A. V.
Chukanov, D. V. Kustov and B. A. Popov. D. N. thanks also V. A. Naumov for
valuable discussions of preliminary versions of this paper.

\end{acknowledgement}
\begin{appendix}
\input{app_su6wf}
\input{app_calc_pol}
\end{appendix}

\input{references}
\newpage
\listoffigures
\newpage
\listoftables
\newpage
\input{all_figures}

\newpage
\input{all_tables}

\end{document}

%% file: intro.tex
\section{Introduction\label{sec:intro}}

The spin stucture of hadrons is still not understood at a fundamental
level in QCD, despite having been extensively studied both theoretically
and experimentally over past two decades. The trigger for many of these
activities was the so-called {\em spin crisis}, namely an unexpectedly
small net quark contribution to the total spin of the nucleon: $0.27 \pm
0.04$ at $Q^2=10$ GeV$^2$, reported initially by the EMC
Collaboration~\cite{EMC}, and subsequently confirmed by a number of other
experiments~\cite{SMC,E143,E155,HERMES}.  Defining $\Delta q$ to be the
net polarization of a given quark species $q$, and likewise $\Delta \bar
q$ for the corresponding antiquarks, it has been found that:

\begin{equation}
\label{eq:deltaq_res}
\begin{tabular}{l}
$\Delta u + \Delta \bar u = \phantom{-}0.82 \pm 0.03$,\\
$\Delta d + \Delta \bar d = -0.44\pm 0.03$,\\
$\Delta s + \Delta \bar s = -0.11 \pm 0.03$.
\end{tabular}
\end{equation}
This result indicates the presence of strange sea quarks
with negative net polarization,
though the interpretation depends strongly on the gluon contribution
to the nucleon spin~\cite{GluonAnomaly}. 
Among the many questions still challenging both theoretical and
experimental efforts we may mention:
\begin{itemize}
\item[---] Are strange quarks in the nucleon indeed polarized as suggested 
by the deep-inelastic scattering (DIS) data?
\item[---] What is the gluon contribution to the nucleon spin?
\item[---] What is the spin content of other baryons?
\end{itemize}
There are several ongoing experimental activities dedicated to
investigating the gluon contribution to the nucleon
spin~\cite{E143,HERMES,compass}.  In principle, it is possible to measure
$\Delta s,\;  \Delta \bar s$ directly in an experiment with neutrino and
anti-neutrino charged-current (CC) and neutral-current (NC) (quasi)
elastic scattering off neutrons and protons~\cite{Alberico:2001sd} and in
DIS from a polarized target~\cite{nufac}. However, for the moment there
are no approved experiments of this type. Therefore, it is of great
interest to have estimates of the polarization of the strange quarks in
the nucleon using indirect measurements.

Using a model of polarized intrinsic strangeness~\cite{polstr} it was
suggested~\cite{ekk} that strange quark polarization could manifest itself
via the polarization of $\lam$ hyperons produced in the target
fragmentation region of lepton-nucleon DIS along an axis that is
longitudinal with respect to the momentum of the exchanged boson. Several
experimental measurements of $\lam$ polarization have been made in
neutrino and anti-neutrino DIS. Longitudinal polarization of $\lam$
hyperons was first observed in bubble chamber (anti) neutrino
experiments~\cite{wa21,wa59,e632}, with the results shown in
Table~\ref{tab:nu-long-pol}. The observed longitudinal polarization of
$\lam$ hyperons varied from $-0.29 \pm 0.18$ to $-0.63 \pm 0.13$,
increasing in absolute value in the region $x_F<0$. However, all
these early experiments suffered from a lack of statistics, and the
results could not be considered conclusive, although they were very
suggestive.

The NOMAD Collaboration has recently published new and interesting
results on $\lam$ and $\alam$ polarization with much larger
statistics~\cite{NOMAD_lambda_polar,NOMAD_alambda_polar,DN_thesis}. The
observed longitudinal polarization of $\lam$ hyperons produced in neutrino
DIS from an isoscalar target was $P_{\lam}=-0.15\pm 0.03(stat.)$, again
increasing in absolute value when $x_F<0$: $P_{\lam}(x_F<0)=-0.21\pm
0.04(stat.)$.

There have been a few attempts in the literature to describe $\lam$ and
$\alam$ polarization in the target fragmentation region. The polarized
intrinsic strangeness model~\cite{ekk} mentioned earlier predicted a
negative sign for the longitudinal polarization of $\lam$ and $\alam$
hyperons. The polarization of $\lam$ hyperons in the target fragmentation
region of DIS has also been considered in the meson cloud model
\cite{melthom}. Due to the pseudoscalar nature of the $N K \lam$ coupling,
the polarization of final-state $\lam$ hyperons was predicted to be
strongly anticorrelated to that of the nucleon, vanishing for an
unpolarized target.

In addition to studies of the nucleon's intrinsic strangeness content, the
self-analyzing properties of $\lam$ and $\alam$ decays are crucial for
investigations of the spin structure of these hadrons in experiments with
a source of polarized fragmenting quarks. The possible correlation of the
polarizations of the scattered quark and the final-state hadron is
described by the spin transfer coefficient:
\begin{equation}
C_q^h(z) \equiv \Delta D_q^h(z)/D_q^h(z),
\end{equation}
where $D_q^h(z)$ and $\Delta D_q^h(z)$ are unpolarized
and polarized fragmentation functions for the quark $q$ to yield a hadron
$h$ with a fraction $z$ of the quark energy. We recall that $\lam$ and
$\alam$ polarization has been, or is planned to be measured in several
different processes where the fragmentation of polarized quarks can be 
measured:
\begin{itemize} 
\item in $e^+e^-$ annihilation at the $Z^0$ pole at LEP~\cite{LEP_LAMBDA}, 
\item in polarized charged-lepton DIS off a target 
nucleon~\cite{compass,hermes,e665}, 
\item in (anti) neutrino DIS off a target
nucleon~\cite{wa21,wa59,e632,NOMAD_lambda_polar,NOMAD_alambda_polar,DN_thesis}.
\end{itemize}
These processes have been extensively studied in different theoretical 
models~\cite{BJ,Anselmino_group,Kotzinian_group,Liang:2001yt,Liang_group,Boros:1998kc,Soffer_group,Vogelsang,ashli},
in attempts to understand the spin structure of final-state hadrons and 
spin effects in the quark fragmentation process.
A key assumption, adopted widely, is that the struck
quark fragmentation can be disentangled from the nucleon remnant
fragmentation by imposing a cut: $x_F>0$. We discuss in 
Sec.~\ref{sec:frag_model} the validity of this
assumption, which is important also for attempts to probe intrinsic 
strangeness.

It is important to recall that a significant fraction of $\lam$ hyperons
are produced via decays of heavier strange particles, such as $\Sigma^0$,
$\Sigma^\star$ and $\Xi$. As was first pointed by Bigi~\cite{bigi}, in
polarized lepton-nucleon DIS these heavy hyperons may inherit polarization
from the remnant diquark left behind when the struck quark is removed, or
from the polarization of the fragmenting quark. The $\lam$ hyperons
originating from decays of these hyperons will then also acquire some
polarization, and this effect should be taken into account in any
theoretical consideration.

We formulate in Section~\ref{sec:model} of this paper a refinement of the
polarized intrinsic strangeness model~\cite{polstr,ekk}, in which $s -
{\bar s}$ pairs are associated with individual valence quarks. The
intrinsic strange quarks and antiquarks may then have polarizations that
are (anti)  correlated with the valence quark polarizations, rather than
with that of the nucleon as a whole. We assume that the valence quark wave
functions are described by non-relativistic $SU(6)$ wave functions. In
Section~\ref{sec:diquark_pol} we use this model to calculate the
polarizations of $\lam$ hyperons produced promptly or via the decays of
heavier hadrons from fragmentation of the the remnant diquark, and in
Section~\ref{sec:quark_pol} we calculate the effect of struck quark
fragmentation. The fragmentation model used in this paper is described in
Section~\ref{sec:frag_model}, and we fix free parameters of our model from
a fit to the NOMAD data. In Section~\ref{sec:results} we present the model
predictions for the polarization of $\lam$ hyperons. Those produced in
(anti) neutrino DIS are discussed in Section~\ref{sec:results_neutrino}
and those in charged-lepton DIS are discussed in
Section~\ref{sec:results_leptons}. Wherever possible, these predictions
are compared with available data.  Finally, in Section~\ref{sec:summary}
we draw our conclusions.

%% file: method.tex
\section{Model for $\lam$ Polarization\label{sec:method}}
\input{model}

\input{tfr}
\input{cfr}

\input{fragmentation_model}

\input{fit}

%% file: model.tex
\subsection{Polarized Intrinsic Strangeness Model\label{sec:model}}

The main idea of the polarized intrinsic strangeness model applied to
semi-inclusive DIS is that the polarization of $s$ quarks and $\bar s$ 
antiquarks in the hidden strangeness component of the nucleon wave 
function should be (anti)correlated with that of the struck quark. This 
correlation is
described by the spin correlation coefficients $C_{sq}$
\begin{equation}
\label{eq:spol}
P_s=C_{sq} P_q,
\end{equation}
where $P_q$ and $P_s$ are the polarizations of the initial struck
(anti)quark and remnant $s$ quark.
  
Such a correlation can be motivated by a strong attraction between quark
and antiquark in the spin-singlet pseudoscalar $J^{P}=0^{-}$ channel, and
the fact that vacuum quark-antiquark pairs must be in a relative
spin-triplet $^3P_0$ state.  In the original version of this
model~\cite{polstr,ekk}, the spin projection of the $\bar{s} {s}$ pair on
the direction of the struck quark spin was taken to be $S_z(\bar s s)=-1$,
{\it i.e.}, maximal anticorrelation between the polarizations of the
struck quark and the remnant $s$ quark was assumed, corresponding to a
spin correlation coefficient $C_{sq}=-1$.  However, other values of
$C_{sq}$ are also conceivable. For example, an alternative scenario for
polarized intrinsic strangeness in which the $S_z(\bar s s)=-1$ and
$S_z(\bar s s)=0$ states of the $\bar{\sf s} {\sf s}$-pair are equally
probable has been considered in~\cite{nufac,akqdq}. In that case, one
would have $C_{sq}=-\frac{1}{3}$. Moreover, the spin correlation
coefficients could in principle be different for scattering off a valence
and a sea quark: ($C_{sq_{val}} \ne C_{sq_{sea}}$).

In this paper we refine the polarized intrinsic strangeness 
model~\cite{polstr,ekk} as follows:
\begin{itemize}
\item[---] We leave $C_{sq_{val}}$ and $C_{sq_{sea}}$ as free
  parameters, that are fixed in a fit to the NOMAD
  data~\cite{NOMAD_lambda_polar}. 
\item[---] We take into account the polarization transfer from 
  heavier hyperons.
\end{itemize}

\subsection{Polarization of Strange Hadrons in Diquark 
Fragmentation\label{sec:diquark_pol}}   

After removing a polarized scattered quark from an unpolarized nucleon,
the remnant diquark may combine with an $s$ quark - which could originate
from the nucleon sea or from a colour string between the diquark and the
scattered quark - to form a strange baryon $Y$.  This baryon could be a
${\lam}$ hyperon - which we term {\em prompt} production - or a heavier
hyperon that decays subsequently into a ${\lam}$.  Polarization transfer
to a promptly-produced ${\lam}$ in $\nu n$ charged-current DIS is shown in
Fig.~\ref{fig:nun}. In general, the strange baryon $Y$ may inherit
polarization form the spin configuration of the remnant diquark and/or the
possible polarization of the strange quark.

We calculate the polarization of ${\lam}$ hyperons assuming the 
combination
of a non-relativistic $SU(6)$ quark-diquark wave function and the
polarized intrinsic strangeness model described above. Specifically, we
use the usual spin-flavor $SU(6)$ wave functions of octet and decuplet
baryons and (\ref{eq:spol}) to model the strange quark polarization.

We define the quantization axis along the three-momentum vector of the
exchanged boson. The longitudinal polarization of ${\lam}$ hyperons
produced via the $l\up N \to l^\prime Y(J,M) X$ process, where the strange
baryon $Y(J,M)$ has total spin $J$ and third component $M$, is given
in the target fragmentation region by:

\begin{eqnarray}
\label{eq:lambda_polar}
&&P_{\lam}^{l q}(Y;N) = \\
&=&P_q \cdot \frac{\sum_M P_s\left(Y(J,M)\right) \mid \langle Y(J,M) \mid 
N\ominus q \rangle \otimes \mid s_{pol}\rangle \mid ^2}
{\sum_M \mid \langle Y(J,M) \mid N\ominus q \rangle \otimes \mid 
s_{pol}\rangle 
\mid ^2},\nonumber
\end{eqnarray}
where $P_q$ is the polarization of the struck quark,
$P_s\left(Y(J,M)\right)$ is the polarization of the strange
quark in the baryon $Y$ with spin state $\mid Y(J,M\rangle)$, and 
$\mid~N\ominus~q~\rangle~\otimes~\mid~s_{pol}~\rangle$ is the combined
wave function of the remnant diquark and 
of the $s$ quark with longitudinal polarization $C_{sq}$.

In (\ref{eq:lambda_polar}), we have taken into account the fact that, in
both the electromagnetic decay of the $\Sigma^0$ and the strong decay of
the $\Sigma^\star$ to ${\lam}$, the non-strange diquark changes its spin
from 1 to 0, while the strange quark retains its
polarization~\cite{ashli}.

%% file: tfr.tex
After some straightforward algebra, we obtain the following predictions of
non-zero polarization for $\Lambda$ hyperons, where more details can be
found in Appendix~\ref{app:details}:

\begin{center}
\begin{eqnarray}
\label{eq:dir}
\nonumber
&&
P_{\Lambda}^{\nu \, d}(prompt;N)=
P_{\Lambda}^{\bar\nu \, u}(prompt;N)= \nonumber\\ 
&&
P_{\Lambda}^{l\, u}(prompt;N)=P_{\Lambda}^{l\, d}(prompt;N)=
C_{sq} \cdot P_q,\nonumber\\
\nonumber\\
&&
P_{\Lambda}^{\nu \, d}(\Sigma^0;n) = 
P_{\Lambda}^{\bar\nu \,u}(\Sigma^0;p) = \nonumber\\ 
&&
P_{\Lambda}^{l \, u}(\Sigma^0;p) =
P_{\Lambda}^{l \, d}(\Sigma^0;n) = 
\frac{1}{3} \cdot \frac{2+C_{sq}}{3+2C_{sq}} \cdot P_q,\nonumber\\
\\
&&
P_{\Lambda}^{\nu \, d}({\Sigma^\star}^0;n) = 
P_{\Lambda}^{\nu \, d}({\Sigma^\star}^+;p) =\nonumber\\[2mm]
&&
P_{\Lambda}^{\bar\nu \, u}({\Sigma^\star}^0;p) =
P_{\Lambda}^{\bar\nu \, u}({\Sigma^\star}^+;n) =\nonumber\\[2mm]
&&
P_{\Lambda}^{l \, u}({\Sigma^\star}^0;p) =
P_{\Lambda}^{l \, d}({\Sigma^\star}^0;n) =\nonumber\\ 
&&
P_{\Lambda}^{l \, d}({\Sigma^\star}^+;p) =
P_{\Lambda}^{l \, u}({\Sigma^\star}^-;n) =
-\frac{5}{3} \cdot \frac{1-C_{sq}}{3-C_{sq}} \cdot P_q\nonumber.
\end{eqnarray}
\end{center}
In these expressions, $l$ stands for either the charged lepton in an
electromagnetic interaction or the (anti)neutrino in a neutral-current
interaction. The interacting quark polarization, denoted by $P_q$ in
(\ref{eq:dir}), is different in different processes, and depends in
general on the quark flavour.  The polarization of the interacting quark
($d$ and $u$) for (anti)neutrino interactions via both charged and neutral
current and for charged-lepton interactions via the electromagnetic
current is shown in Table~\ref{tab:pq}. In this Table, $\xi \equiv
2/3\sin^2\theta_W$, $P_B$ is the charged-lepton longitudinal polarization,
and $D(y) = \frac{1-(1-y)^2}{1+(1-y)^2}$ is the virtual photon
depolarization factor, while $y$ represents the relative energy loss of
the lepton.

%% file: cfr.tex
\subsection{Polarization of Strange Hadrons in Quark Fragmentation
\label{sec:quark_pol}}

We now discuss in more detail how strange hadrons produced in the 
fragmentation of a polarized quark could
also be polarized. The polarization of ${\lam}$ hyperons produced promptly
or via a strange baryon $Y$ in quark fragmentation is related to the quark
polarization $P_q$, see Table~\ref{tab:pq}, by:

\begin{equation}
\label{eq:cfr}
P_{\lam}^q(Y) = - C_q^{\lam}(Y) P_q,
\end{equation}
where $C_q^{\lam}(Y)$ is the corresponding spin transfer coefficient. The
negative sign in (\ref{eq:cfr}) appears because of the opposite
polarizations of the {\em incoming} and {\em outgoing} quarks: see
Fig.~\ref{fig:nun}.

We use two different models to calculate $C_q^{\lam}(Y)$. The first one is
based on non-relativistic $SU(6)$ wave functions, where the ${\lam}$ spin
is carried only by its constituent $s$ quark. In this case, the
promptly-produced ${\lam}$ hyperons could be polarized only in $s$ quark
fragmentation. However, ${\lam}$ hyperons produced via the decays of
heavier strange hyperons originating from $u$ or $d$ quark fragmentation
could also be polarized.

The second approach was suggested by Burkardt and Jaffe~\cite{BJ}, who
assumed that the `spin crisis' exists not only for the nucleon, but also
for other octet hyperons. Then, using flavor $SU(3)$ symmetry and
polarized DIS data, they concluded that the $u$ and $d$ quarks inside the
${\lam}$ hyperon are polarized in the opposite direction to the ${\lam}$
spin at the level of $-20\%$ each. In the same way, the spin contents of
all octet baryons were obtained in~\cite{Boros:1998kc}. Then, they assumed
that the polarized quark-to-hyperon fragmentation function was equal to
the quark distribution function in that hyperon. We note that, in this
Burkardt-Jaffe (BJ) model, all the strong-interaction effects are in
principle included in the fragmentation and distribution functions.
Therefore, one should not consider spin transfer from $\Sigma^*$
hyperons to the ${\lam}$ hyperon in BJ model. Table~\ref{tab:cfr_coeff}
summarizes the spin correlation coefficients in the $SU(6)$ and BJ models
for both prompt ${\lam}$ hyperons and octet and decuplet intermediate
hyperons.

%% file: fragmentation_model.tex
\subsection{Fragmentation Models\label{sec:frag_model}}

In order to apply our formalism to a real experimental environment,
it is important to know the relative fractions of $\lam$
hyperons produced in different channels in both quark and diquark
fragmentation in different regions of the DIS phase space. 

It is well known from inclusive reactions at high energies that the
typical hadronic rapidity correlation length is $\Delta y_h \simeq 2$.
Thus to select, for example, the current fragmentation region one has to
choose the hadrons with centre-of-mass rapidity
$y_h^{CM}>2$~\cite{berger}. According to this criterion, the minimal
fraction $z_{min}$ of the energy transferred to the struck quark required
to select particles produced in the quark fragmentation region depends on
the particle species and on the invariant mass of hadronic system, $W$.
For example~\cite{mulders}, in the case of $\lam$ production at $W=5$ GeV,
$\Delta y_h < 1.5$ at any value of $z$, and at $W=20$ GeV the criterion is
satisfied only for $z > 0.5$, where the production yield of $\lam$
hyperons is strongly suppressed.

{\it However, as we now show, the beam energies in current experiments on
$\lam$
polarization~\cite{compass,NOMAD_lambda_polar,NOMAD_alambda_polar,hermes,e665}
are not enough for the $x_F>0$ region to be populated by $q \to \lam$
fragmentation only}.

To describe $\lam$ production and polarization in the full $x_F$ interval,
we use the LUND string fragmentation model, as incorporated into the {\tt
JETSET7.4} program~\cite{JETSET}. We use the {\tt LEPTO6.5.1}~\cite{lepto}
Monte Carlo event generator to simulate charged-lepton and (anti)neutrino
DIS processes. With suitable choices of input parameters, these event
generators reproduce well the distributions of various observables and
particle yields. It was recently found by the NOMAD Collaboration that
{\tt JETSET7.4} with its default set of parameters overestimates the
observable yields of strange hadrons produced in neutrino charged-current
interactions by factors of 1.5 to 3~\cite{NOMAD_resonances}. Therefore, we
use here the new set of {\tt JETSET} parameters found by the NOMAD
Collaboration~\cite{NOMAD_JETSET}, so as to adjust the yields of strange
particles in DIS processes.  We have also modified the {\tt LEPTO} code as
follows:
\begin{itemize}
\item[--] The cut on the minimal value of the energy in a fragmenting
  jet is removed. This was found by the NOMAD Collaboration to 
  reproduce better the observed $W^2$ distribution in neutrino 
charged-current events.  
\item[--] A split of the nucleon is implemented when the 
  exchanged boson interacts with a sea $u$ or $d$ quark. 
In the default version of the {\tt LEPTO} code, this is done for all sea 
quarks except $u$ and $d$ sea quarks. 
\end{itemize} 

In the framework of {\tt JETSET}, it is possible to trace the particles'
parentage. We use this information to check the origins of the strange
hyperons produced in different kinematic domains, especially at various
$x_F$. It is widely assumed that the current fragmentation region, defined
to have $x_F>0$, corresponds to the quark fragmentation region. However,
as we mentioned above, arguing on the basis of the
criterion~\cite{berger}, this is not true at moderate energies, even for
high $z$ $\lam$ hyperons. According to the {\tt LEPTO} and {\tt JETSET}
event generators, the $x_F$ distribution of the diquark to $\lam$
fragmentation is weighted towards large negative $x_F$. However, its tail
in the $x_F>0$ region overwhelms the quark to $\lam$ $x_F$ distribution at
these beam energies. In Fig.~\ref{fig:xf_default}, we show the $x_F$
distributions of $\lam$ hyperons produced in diquark and quark
fragmentation, as well as the final $x_F$ distributions. We tag $\lam$
hyperons as being produced by quark and diquark fragmentation according to
the rank of these particles in two different descriptions called `model A'
and `model B', which we explain in Section~\ref{sec:fit_to_nomad}. The
distributions in Fig.~\ref{fig:xf_default} are shown for $\nu_\mu$ CC DIS
at the NOMAD mean neutrino energy $E_\nu = 43.8$~GeV, and for $\mu^+$ DIS
at the COMPASS muon beam energy $E_\mu = 160$~GeV. The relatively small
fraction of the $\lam$ hyperons produced by quark fragmentation in the
region $x_F > 0$ is related to the relatively small centre-of-mass
energies - about 3.6 GeV for HERMES, about 4.5 GeV for NOMAD, about 8.7
GeV for COMPASS, and about 15 GeV for the E665 experiment - which
correspond to low $W$. The authors~\cite{Liang:2001yt}, using on Lund and
{\tt JETSET} generators, also reported that the mean $W^2$ in HERMES and
NOMAD is too low to populate the $x_F>0$ region with struck-quark
fragmentation alone.


The situation changes drastically for (anti)neutrino experiments, if the
beam energy is increased to 500 GeV. As one can see in the left part of
Fig.~\ref{fig:xf_higher_energy}, now the $x_F>0$ region is populated
mainly by $\lam$ hyperons produced in quark fragmentation. However, the
beam energy would have to be further doubled in the charged-lepton DIS
process, in order to populate the $x_F>0$ region by struck-quark
fragmentation. This is related to the lower $\langle W\rangle$ for
electromagnetic interactions compared to that in (anti)neutrino DIS at the
same beam energies.

%% file: fit.tex
\subsection{Fixing Free Parameters of the Model.\label{sec:fit_to_nomad}}

We now compute the $\Lambda$ polarization produced in any allowed 
kinematic domain, taking into account
the $\Lambda$ origins predicted by the LUND model, using the 
formalism developed in 
Sections~\ref{sec:diquark_pol} and \ref{sec:quark_pol}. We assume that 
$\lam$ hyperons could be polarized only 
if they have one of the following origins:
\begin{enumerate} 
\item Prompt $\lam$ hyperons containing the struck quark,
\item $\lam$ hyperons produced in the decays of heavier strange hyperons 
containing the struck quark,
\item Prompt $\lam$ hyperons containing the target nucleon remnant,
\item $\lam$ hyperons produced in the decays of heavier strange hyperons 
containing the remnant diquark.
\end{enumerate}

In the framework of the {\tt JETSET} generator, it is possible to identify
$\lam$ hyperons unambiguously in one of these categories, since all
hadrons generated in the Lund colour-string fragmentation model are
ordered in rank from one end of the string to the other. Therefore, we
introduce two rank counters: $R_{qq}$ and $R_q$ which correspond to the
particle rank from the diquark and quark ends of the string,
correspondingly. A hadron with $R_{qq}=1$ or $R_q=1$ would contain the
diquark or the quark from one of the ends of the string.

However, one should perhaps not rely too heavily on the tagging specified
in the LUND model. Therefore, we consider the following two variant
fragmentation models:

{\bf Model A:}
\begin{itemize}
\item The hyperon contains the stuck quark only if $R_q=1$, and
\item The hyperon contains the remnant diquark only if $R_{qq}=1$.
\end{itemize}

{\bf Model B:}
\begin{itemize}
\item The hyperon contains the stuck quark if $R_q\ge 1$ and $R_{qq} \ne 
1$, and
\item The hyperon contains the remnant diquark if $R_{qq}\ge 1$ and 
$R_q\ne 1$.
\end{itemize}
Clearly, Model B weakens the Lund tagging criterion by averaging over the 
string, whilst retaining information on the end
of the string where the hadron originated.

We vary the two correlation coefficients $C_{sq_{val}}$ and $C_{sq_{sea}}$
in fitting Models A and B to the NOMAD $\Lambda$ polarization
data~\cite{NOMAD_lambda_polar}. Scattering of the exchange $W$ boson on
the sea quark is expected to be enhanced in $\lam$ hyperons produced at
low $x_{Bj}$ or high $W^2$. A strong target nucleon effect was also found
by NOMAD. Therefore, we fit the following 4 NOMAD points to determine our
free parameters:

\begin{itemize}
\item $\nu p$: $P_x^\Lambda = -0.26 \pm 0.05(stat)$,
\item $\nu n$: $P_x^\Lambda = -0.09 \pm 0.04(stat)$,
\item $W^2<15$ GeV$^2$: $P_x^\Lambda(W^2<15) = -0.34 \pm 0.06(stat)$,
\item $W^2>15$ GeV$^2$: $P_x^\Lambda(W^2>15) = -0.06 \pm 0.04(stat)$.
\end{itemize}
We find from these fits similar values for both the $SU(6)$ and BJ models:
\begin{itemize}
\item[] {\bf Model A:} $C_{sq_{val}} = -0.35 \pm 0.05$, $C_{sq_{sea}} = 
-0.95 \pm 0.05$. 
\item[] {\bf Model B:} $C_{sq_{val}} = -0.25 \pm 0.05$, $C_{sq_{sea}} = 
0.15 \pm 0.05$. 
\end{itemize}
The error bars are obtained from the variation in the $\chi^2$ functional 
around its minimum. 

The coefficients found in models A and B are quite different, as could be
anticipated from the different descriptions of the strange hadrons'
origins in these models. However, the two fragmentation models give
similar predictions for the polarization of $\lam$ hyperons produced in
various lepton-nucleon DIS experiments, as we discuss in the next Section.

%% file: results.tex
\section{Results and Discussion\label{sec:results}}

We present in this Section our model predictions for the longitudinal
polarization of $\lam$ hyperons as a function of the kinematic variables
$W^2$, $Q^2$, $x_{Bj}$, $y_{Bj}$, $x_F$ and $z$ in both charged- and
neutral-current DIS for neutrinos and antineutrinos, and in charged-lepton
DIS processes. Where data are available, we compare them with our model
predictions.

\subsection{$\Lambda$ Polarization in (Anti)Neutrino 
DIS\label{sec:results_neutrino}}

\subsubsection{Charged Currents}

Our model predictions for the polarization of $\Lambda$ hyperons produced
in $\nu_\mu$ charged-current DIS interactions off nuclei as functions of
different kinematic variables are shown in Fig.~\ref{fig:numuxcc_6}. The
predictions are compared with the observed dependences of the $\lam$
polarization found by the NOMAD
Collaboration~\cite{NOMAD_lambda_polar,DN_thesis}. We see that the
predictions of models A and B are both in quite good agreement with the
NOMAD data, with model B perhaps being slightly preferred by the $W^2$ and
$x_{Bj}$ distributions. We note that $\lam$ polarization in quark
fragmentation is calculated using the $SU(6)$ model for the spin transfer
- see Table~\ref{tab:cfr_coeff}. However, as mentioned in
Section~\ref{sec:frag_model}, the fraction of $\lam$ hyperons produced via
quark fragmentation is relatively small even at $x_F>0$. Thus the
predictions of the $SU(6)$ and BJ~\cite{BJ} spin-transfer models for the
polarization of $\lam$ hyperons are indistinguishable within the
experimental errors, at the present energies of DIS on fixed targets, and
both models predict correctly the sign of $\lam$ polarization.

The NOMAD Collaboration has measured separately the polarization of $\lam$
hyperons produced off proton and neutron targets. Our model predictions
are compared to the NOMAD data in Table~\ref{tab:numuxcc}. We observe good
agreement, within the statistical errors, between the model B description
and the NOMAD data. On the other hand, although model A reproduces quite
well the polarization of $\lam$ hyperons produced from an isoscalar
target, it does not describe so well the separate proton and neutron data.


The predictions for $\bar\nu_\mu$ charged-current DIS are shown in
Fig.~\ref{fig:anumucc_6}, for which data are currently not available. The
neutrino energy was taken to be $E_\nu=43.8$ GeV, which corresponds to the
mean neutrino energy in the NOMAD experiment for events containing
identified $\lam$ hyperons. The predicted dependence on the target nucleon
is summarized in Table~\ref{tab:anumucc}.

\subsubsection{Neutral Currents}


The degree of $\lam$ polarization in (anti)neutrino neutral-current DIS
processes is of great interest, since the $Z^0$ boson interacts with both
flavours of valence quark, in contrast to charged-current interactions.
Therefore, it is possible in principle to check the universality of
$C_{sq}$ coefficients using neutral-current data. We have been informed
that the NOMAD Collaboration plans to investigate for the first time the
polarization of $\lam$ and $\alam$ hyperons produced in neutrino
neutral-current DIS. Also, we observe that this process is of potential
interest for a future neutrino factory~\cite{nufac}.

Therefore, we provide here our model predictions for the polarization of
$\Lambda$ hyperons as a function of $W^2$, $Q^2$, $x_{Bj}$, $y_{Bj}$,
$x_F$ and $z$ produced in the neutral-current DIS interactions with nuclei
of neutrinos, in Figs.~\ref{fig:numuxnc_6}, and antineutrinos, in
Figs.~\ref{fig:anumunc_6}. In Tables~\ref{tab:numuxnc} and
\ref{tab:anumunc} we summarize our model predictions for the target
nucleon effects for $\nu_\mu$ and $\bar\nu_\mu$ NC DIS processes
respectively. The beam energy is again taken to be 43.8 GeV.


\subsection{$\Lambda$ Polarization and Spin Transfer in 
Charged-Lepton-Nucleon
DIS\label{sec:results_leptons}}

The sign of the polarization of $\Lambda$ hyperons produced in polarized
charged-lepton DIS off and unpolarized nucleon depends on the sign of the
beam polarization. Therefore we provide our model predictions for the spin
transfer, which is defined as ${P_\Lambda}/{P_B D(y)}$. We note that the
spin transfer is positive for negatively-polarized $\lam$ hyperons
produced in the scattering of negatively-polarized beams.
 
\subsubsection{HERMES}

The HERMES experiment at HERA is collecting data on the polarization of
$\lam$ hyperons produced in polarized $e^+$ beam interactions with
different targets, including protons and neon nuclei~\cite{hermes}. We
present here our model predictions for the spin transfer of $\lam$
hyperons, imposing the kinematic criteria used by HERMES Collaboration.  
The spin transfer as a function of kinematic variables predicted by our
model is compared with the available HERMES data in
Fig.~\ref{fig:e+xxxem_6}. The observed $x_F$ and $z$ dependences of the
$\lam$ spin transfer are compatible with our predictions. An investigation
of the dependences of the $\lam$ polarization on other kinematic variables
could be an important check of our model. More statistics would be needed
to make a detailed comparison, and these may soon be provided by the new
runs of HERA.




\subsubsection{E665}

The E665 Collaboration has reported~\cite{e665} a measurement of the spin 
transfer to
$\lam$ and $\alam$ hyperons produced by the electromagnetic DIS of 470 GeV
$\mu^+$. The kinematic domain used for the measurement was:  
$x_{Bj}<0.1$, $0.25 \mbox{ GeV}^2 < Q^2 < 2.5 \mbox{ GeV}^2$.  In
Fig.~\ref{fig:e665_mu+xxem_6} we present our model predictions for the
spin transfer to $\Lambda$ hyperons produced in $\mu^+$ DIS interactions
on nuclei in this kinematic domain. Unfortunately, the large experimental
errors render the comparison with our model predictions inconclusive.


\subsubsection{COMPASS}


The COMPASS Collaboration plans to investigate the polarization of $\lam$
hyperons produced in the DIS of polarized $\mu^+$ on a $^6Li D$ target.
The beam energy and polarization are 160~GeV and -0.8, respectively.

The calculated value of the spin transfer for events with with $Q^2>1$
GeV$^2$ is presented as a function of different kinematic variables in
Fig.~\ref{fig:compass_mu+xxem_6}. Thanks to the large statistics expected
in this experiment, one can select kinematic regions where the predicted
polarization is very sensitive to the value of the spin correlation
coefficient for sea quarks, $C_{sq_{sea}}$. For example, in the region
$x_F>-0.2$, which is experimentally accessible, and imposing the cut
$0.5<y<0.9$, one ensure a large spin transfer from the incident lepton to
the struck quark, and enhance the contribution from the sea quarks. The
predicted $\Lambda$ polarization is presented in Table~\ref{tab:compass}.

%% file: conclusion.tex
\section{Summary\label{sec:summary}}

We now summarize the key points of this paper. We have developed a model
for $\Lambda$ polarization in DIS that combines polarized intrinsic
strangeness with non-relativistic $SU(6)$ wave functions. This model has
been combined with the Lund model to describe the fragmentation processes.
We have emphasized that the struck-quark and target fragmentation regions
overlap significantly in experiments in the energy range currently
available, and this effect is taken into account in our calculations.

Our model has only two free parameters, which are fixed from a fit to
NOMAD data on the longitudinal polarization of $\lam$ hyperons produced in
neutrino DIS. Our model describes well the various data available from
this experiment~\cite{NOMAD_lambda_polar,DN_thesis} and from experiments
on charged-lepton DIS~\cite{hermes,e665}. We have proposed two model
variants that differ in the extent to which Lund tagging information is
used, and variant B is favoured by NOMAD data in which DIS off proton
and neutron targets are separated. We have presented predictions for
future data, including neutral-current DIS from NOMAD and charged-lepton
DIS from HERMES and COMPASS. The dependences on the model details
predicted here provide many possibilities for further checks of our
approach.


%% file: app_su6wf.tex
\section{- $SU(6)$ Wave Functions of Octet and Decuplet 
Hyperons\label{app:app_su6wf}}

The quark-diquark $SU(6)$ wave functions of octet and decuplet baryons  
for non-vanishing matrix elements in (\ref{eq:lambda_polar}) read:

\begin{eqnarray}
  \label{eq:baryons}
  \Lambda\up &=& \frac{1}{\sqrt{12}}
  [2(ud)_{0,0}\ups  + \sqrt{2}(us)_{1,1}\downd - \nonumber\\
  && -((us)_{1,0} - (us)_{0,0})\upd - \sqrt{2}(ds)_{1,1}\downu \nonumber\\
  && + ((ds)_{1,0} - (ds)_{0,0})\upu
  ],\nonumber\\
  {\Sigma^0}\up &=& \frac{1}{\sqrt{18}}
  [2(ud)_{1,1}\downs  - \sqrt{2}(ud)_{1,0}\ups +  \nonumber\\
  && + \frac{1}{\sqrt{2}}((ds)_{1,0}+3(ds)_{0,0})\upu -
  (ds)_{1,1}\downu + \nonumber\\
  &&  +  \frac{1}{\sqrt{2}}((us)_{1,0}-3(us)_{0,0})\upd 
  -(us)_{1,1}\downd],\nonumber\\
  {{\Sigma^\star}^+}^\Uparrow &=&
  \frac{1}{\sqrt{3}}[ \sqrt{2}(us)_{1,1}\upu + (uu)_{1,1}\ups]\\
  {{\Sigma^\star}^+}\up &=& \frac{1}{\sqrt{9}}
  [2(us)_{1,0}\upu + \sqrt{2}(us)_{1,1}\downu + \sqrt{2}(uu)_{1,0}\ups
  +  \nonumber\\
  && + (uu)_{1,1}\downs]\nonumber\\
  {{\Sigma^\star}^0}^\Uparrow &=&
  \frac{1}{\sqrt{3}}[(ds)_{1,1}\upu + (ud)_{1,1}\ups + 
  (us)_{1,1}\upd]\nonumber\\
  {{\Sigma^\star}^0}\up &=&
  \frac{1}{\sqrt{18}}
  [2(ds)_{1,0}\upu + \sqrt{2}(ds)_{1,1}\downu + 2(us)_{1,0}\upd\nonumber\\
  && + \sqrt{2}(us)_{1,1}\downd + 2(ud)_{1,0}\ups + \sqrt{2}(ud)_{1,1}\downs    
  ],\nonumber
\end{eqnarray}
where ${\Sigma^\star}^\Uparrow$ (${\Sigma^\star}^\uparrow$)  denotes a
$\Sigma^\star$ state with $M=\frac{3}{2}$ ($M=\frac{1}{2}$). The symbols
$(qq\prime)_{J,M}$ represent the diquark in the spin state $\mid
J,M\rangle$.

%% file: app_calc_pol.tex
\section{- Details of $P_\Lambda$ Calculations\label{app:details}}

We use non-relativistic $SU(6)$ wave functions (see 
App.~\ref{app:app_su6wf}) 
in order to compute the matrix elements in (\ref{eq:lambda_polar}).
We consider here in more detail calculations for $\Lambda$ 
hyperons produced in charged-lepton-nucleon, as the most 
general example.
In this case, the struck quark can be polarized {\em only } if the beam 
or/and target are
polarized. We consider the case of polarized beam and unpolarized target.
After removing a polarized quark $q$ from the nucleon, the product of the 
wave functions of the diquark remnant and $s$ quark reads:

\begin{eqnarray}
\label{eq:lep_remn}
\nonumber
&&\mid N\ominus q_{pol} \rangle \otimes \mid s_{pol} \rangle = \\
&&\frac{1}{2} \left[ \sqrt{1+P_q} \mid N\ominus \upq\rangle 
\left(\sqrt{1+C_{sq}}\ups + \sqrt{1-C_{sq}}\downs\right)+ \right. \nonumber\\
&&+\left.\quad \sqrt{1-P_q} \mid N\ominus \downq\rangle 
\left(\sqrt{1-C_{sq}}\ups+\sqrt{1+C_{sq}}\downs\right) \right], \nonumber
\end{eqnarray}
where $\left(\sqrt{1+C_{sq}}\ups + \sqrt{1-C_{sq}}\downs\right)$
corresponds to the spin part of the polarized $s$ quark wave function, and
$|N\ominus q_{pol} \rangle$ is the wave function of the nucleon
remnant. Explicitly:

\begin{eqnarray}
\label{eq:remn}
&&|p\ominus \upd \rangle =\frac{1}{\sqrt{36}} [-\sqrt{2} (uu)_{1,0} 
+ 2 (uu)_{1,-1}],\nonumber\\
&&|p\ominus \upu \rangle = \frac{1}{\sqrt{36}} [3(ud)_{0,0} 
+ (ud)_{1,0} - \sqrt{2} (ud)_{1,-1}],\nonumber\\[-3mm]
\\
&&|n\ominus \upd \rangle =\frac{1}{\sqrt{36}} [3(ud)_{0,0} 
+ (ud)_{1,0} - \sqrt{2} (ud)_{1,-1}],\nonumber\\
&&|n\ominus \upu \rangle =\frac{1}{\sqrt{36}} [-\sqrt{2} (uu)_{1,0} 
+ 2 (uu)_{1,-1}]\nonumber.
\end{eqnarray}
Using (\ref{eq:lambda_polar}, \ref{eq:baryons}, \ref{eq:remn}), one
can obtain the results presented in (\ref{eq:dir}).

%% file: all_figures.tex
\begin{figure}[htb]
\begin{center}
\epsfig{file=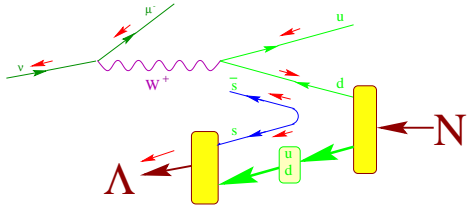,width=0.7\linewidth}
\caption{\label{fig:nun} Polarization transfer to promptly-produced
${\lam}$ hyperons in $\nu n$ charged-current DIS.} 
\end{center}
\end{figure}

\begin{figure}[htb]
\begin{center}
\begin{tabular}{cc}
\epsfig{file=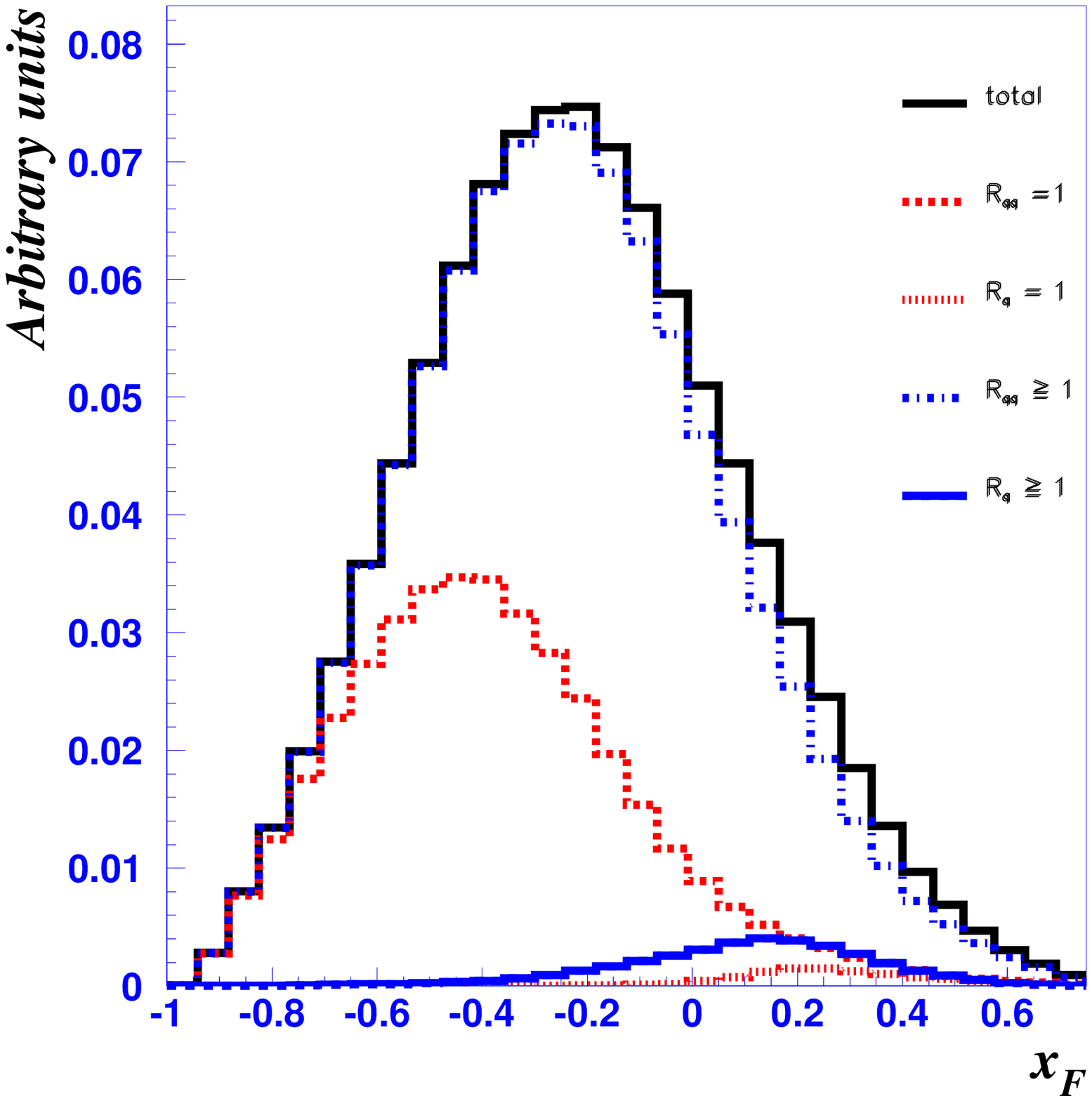,width=0.5\linewidth}&
\epsfig{file=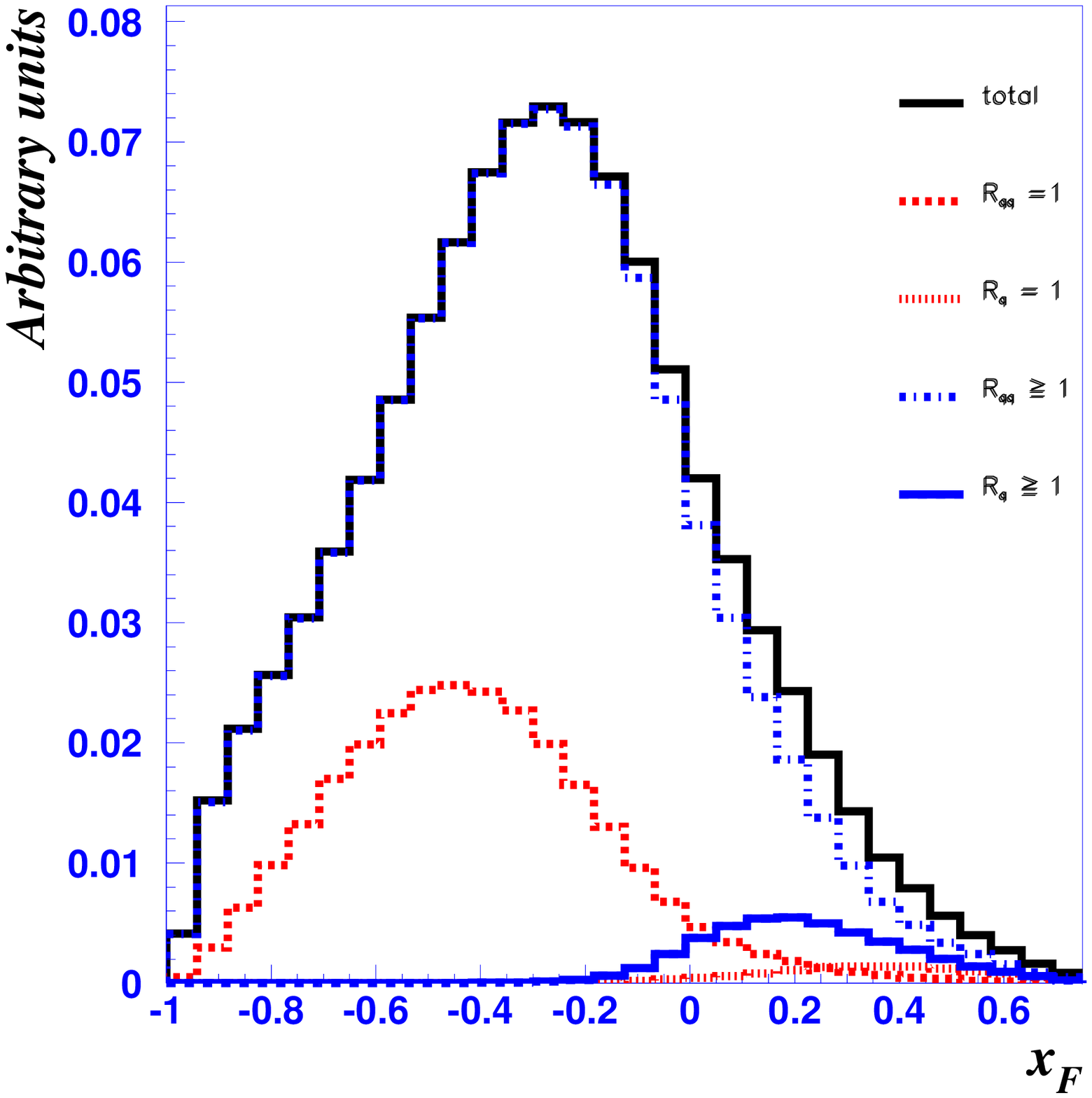,width=0.5\linewidth}\\
\end{tabular}
\caption{\label{fig:xf_default} 
Predictions for the
$x_F$ distributions of all $\lam$ hyperons (solid line), 
of those originating from diquark fragmentation and 
of those originating from quark fragmentation, for the two model variants 
A and B, as explained in the legend on the plots.
The left panel is for $\nu_\mu$ CC DIS with $E_\nu = 43.8$ GeV, 
and the right panel for $\mu^+$ DIS with $E_\mu = 160$ GeV. 
}
\end{center}
\end{figure}

\begin{figure}[htb]
\begin{center}
\begin{tabular}{cc}
\epsfig{file=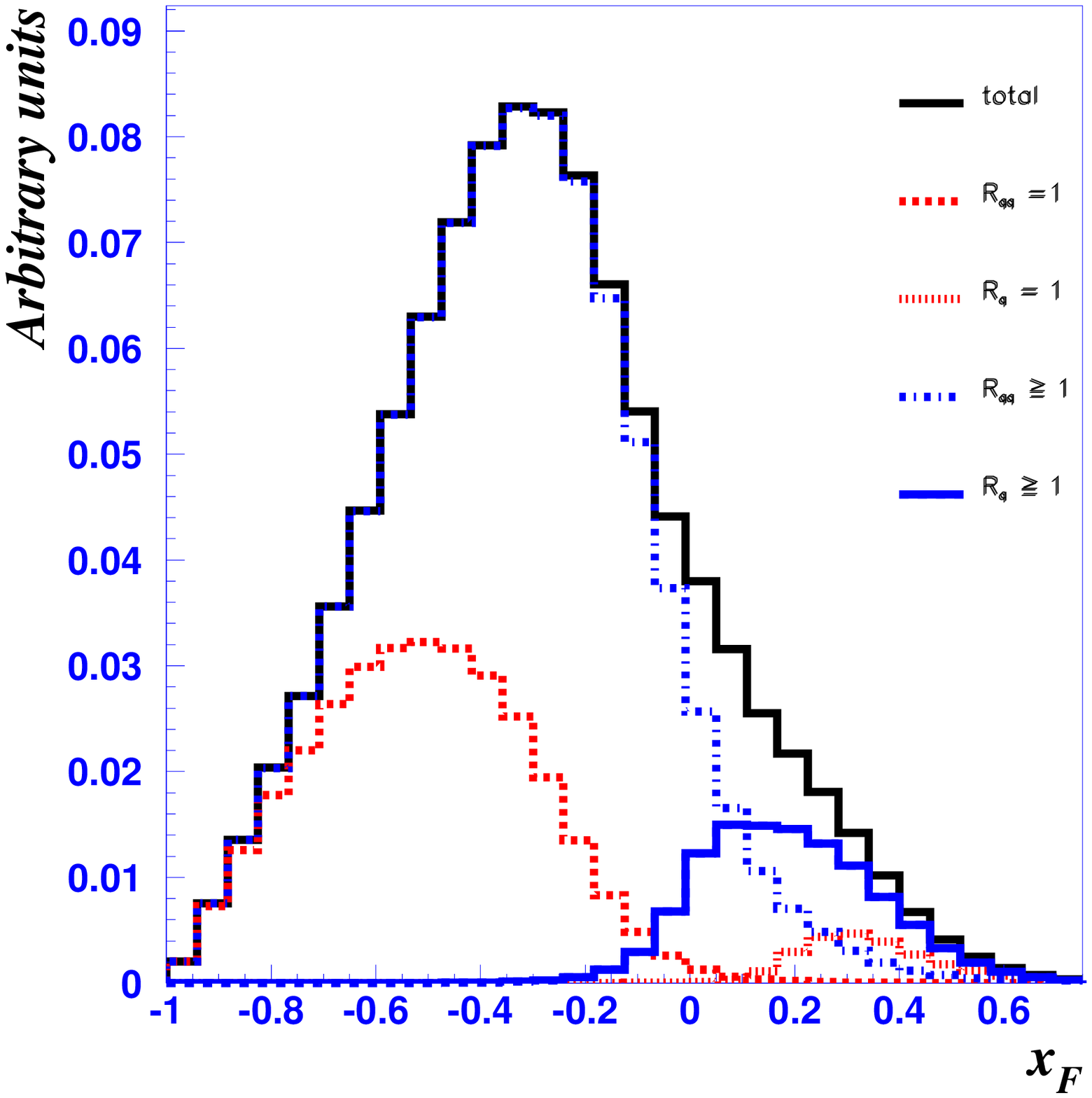,width=0.5\linewidth}&
\epsfig{file=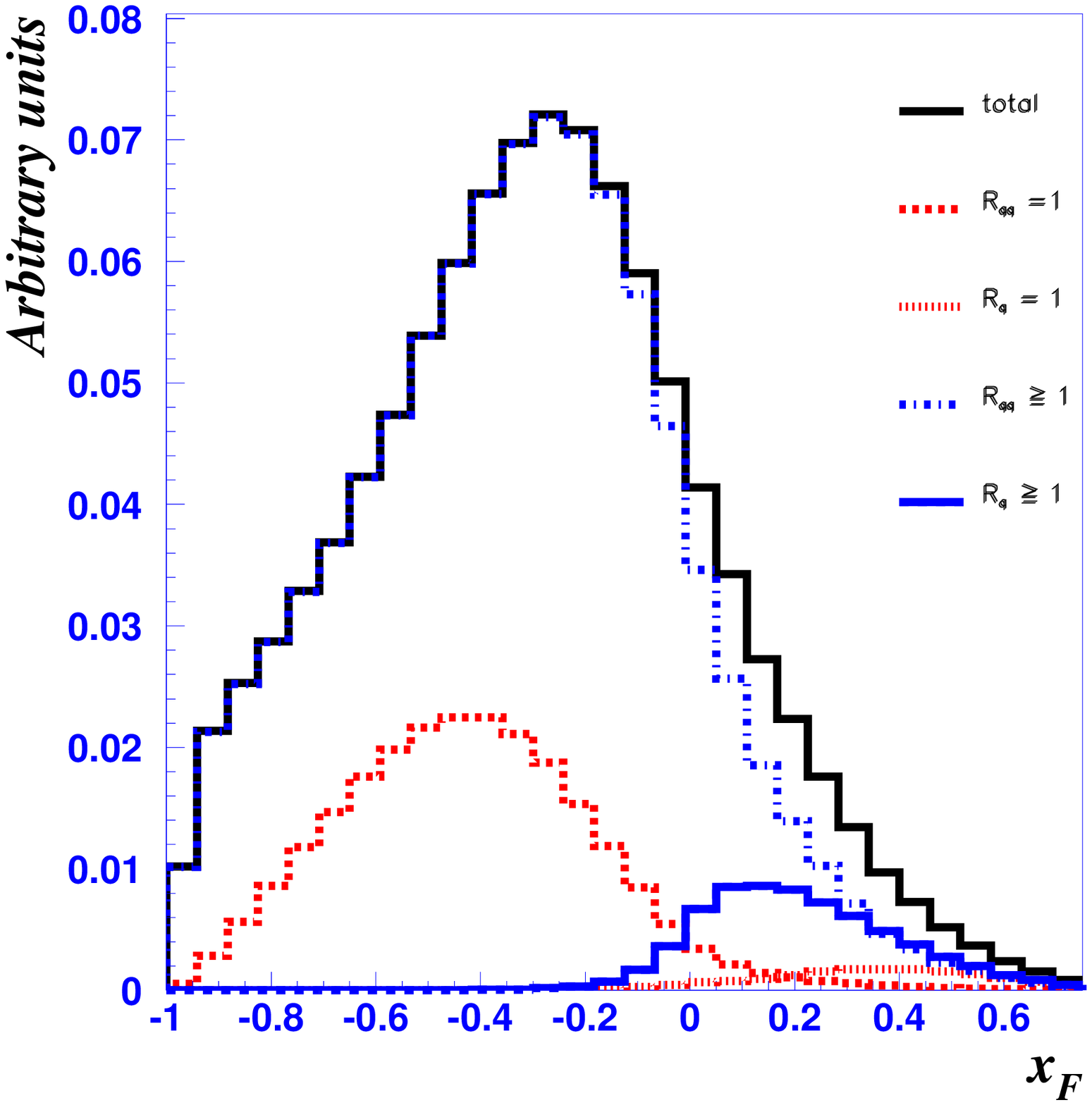,width=0.5\linewidth}\\
\end{tabular}
\caption{\label{fig:xf_higher_energy} 
Predictions for the
$x_F$ distribution of all $\lam$ hyperons (solid line), 
of those originating from diquark fragmentation and 
of those originating from quark fragmentation, for the two model variants 
A and B, as explained in the legend on the plots.
The left panel is for $\nu_\mu$ CC DIS with $E_\nu = 500$ GeV, and the
right panel for $\mu^+$ DIS with $E_\mu = 500$ GeV. 
}
\end{center}
\end{figure}

\begin{figure}[htb]
\begin{center}
\epsfig{file=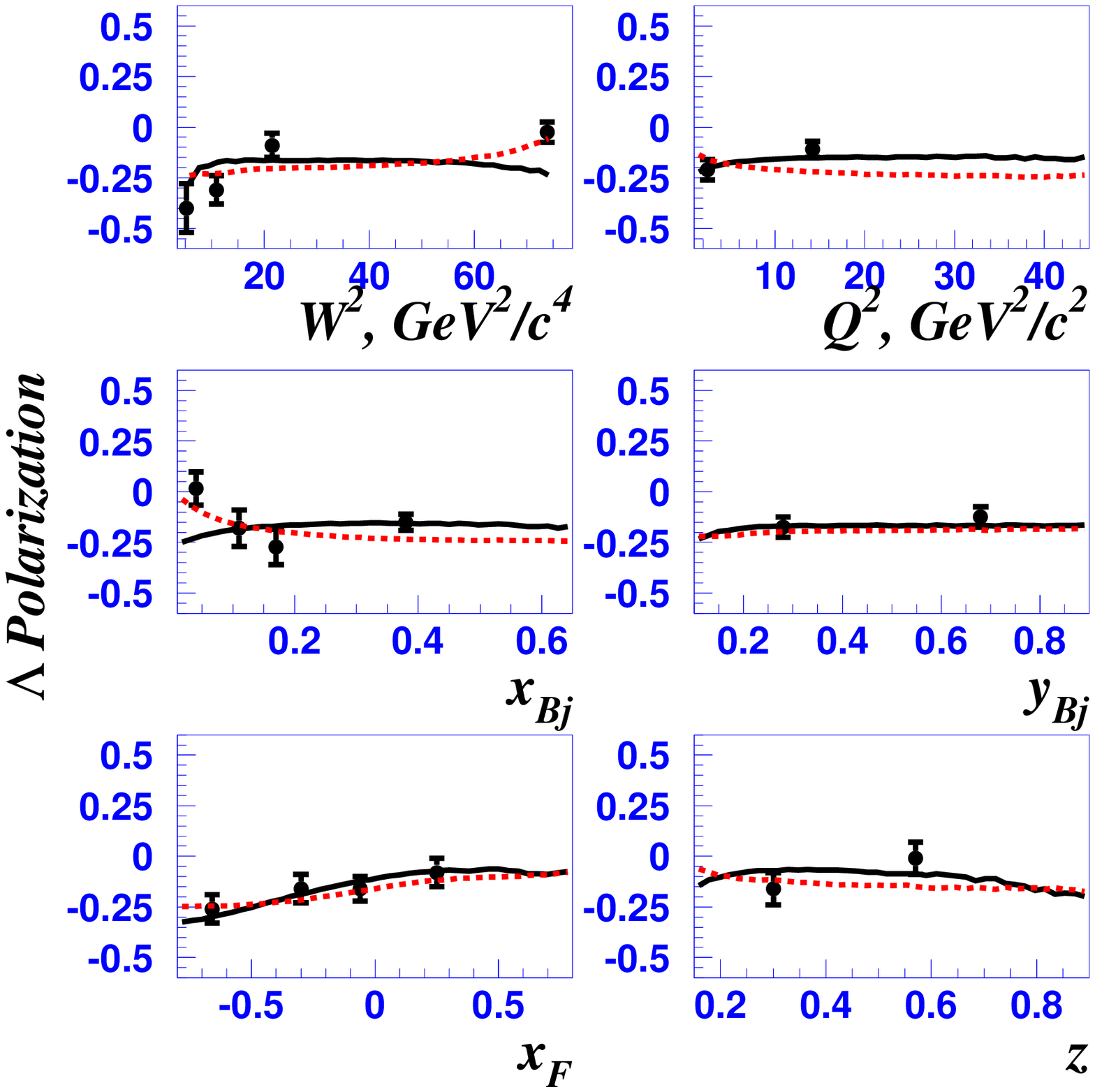,width=\linewidth}
\caption{\label{fig:numuxcc_6} The predictions of model A - solid line 
and model B - dashed line, for the polarization
of $\Lambda$ hyperons produced in $\nu_\mu$ charged-current DIS
interactions off nuclei as functions of $W^2$, $Q^2$, $x_{Bj}$,
$y_{Bj}$, $x_F$ and $z$ (at $x_F>0$). The points with error bars are
from ~\cite{NOMAD_lambda_polar,DN_thesis}.} 
\end{center}
\end{figure}

\begin{figure}[htb]
\begin{center}
\epsfig{file=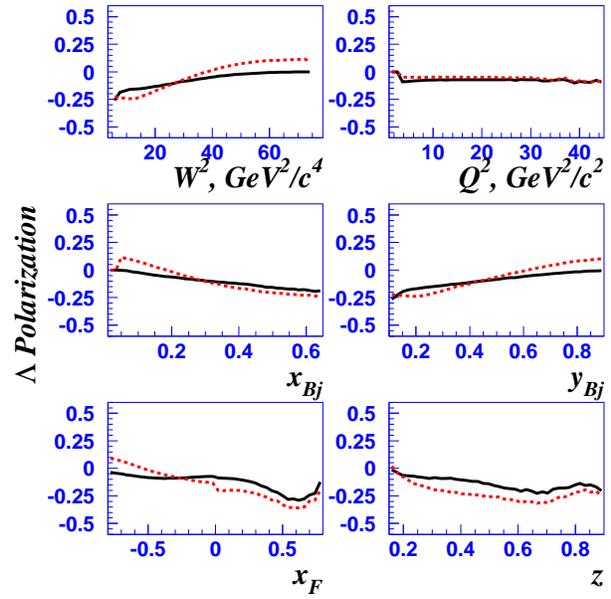,width=\linewidth}
\caption{\label{fig:anumucc_6} The predictions of model A - solid 
line and model B - dashed line, for the polarization
of $\Lambda$ hyperons produced in $\bar\nu_\mu$ charged-current DIS
interactions off nuclei as functions of $W^2$, $Q^2$, $x_{Bj}$ and
$y_{Bj}$.} 
\end{center}
\end{figure}

\begin{figure}[htb]
\begin{center}
\epsfig{file=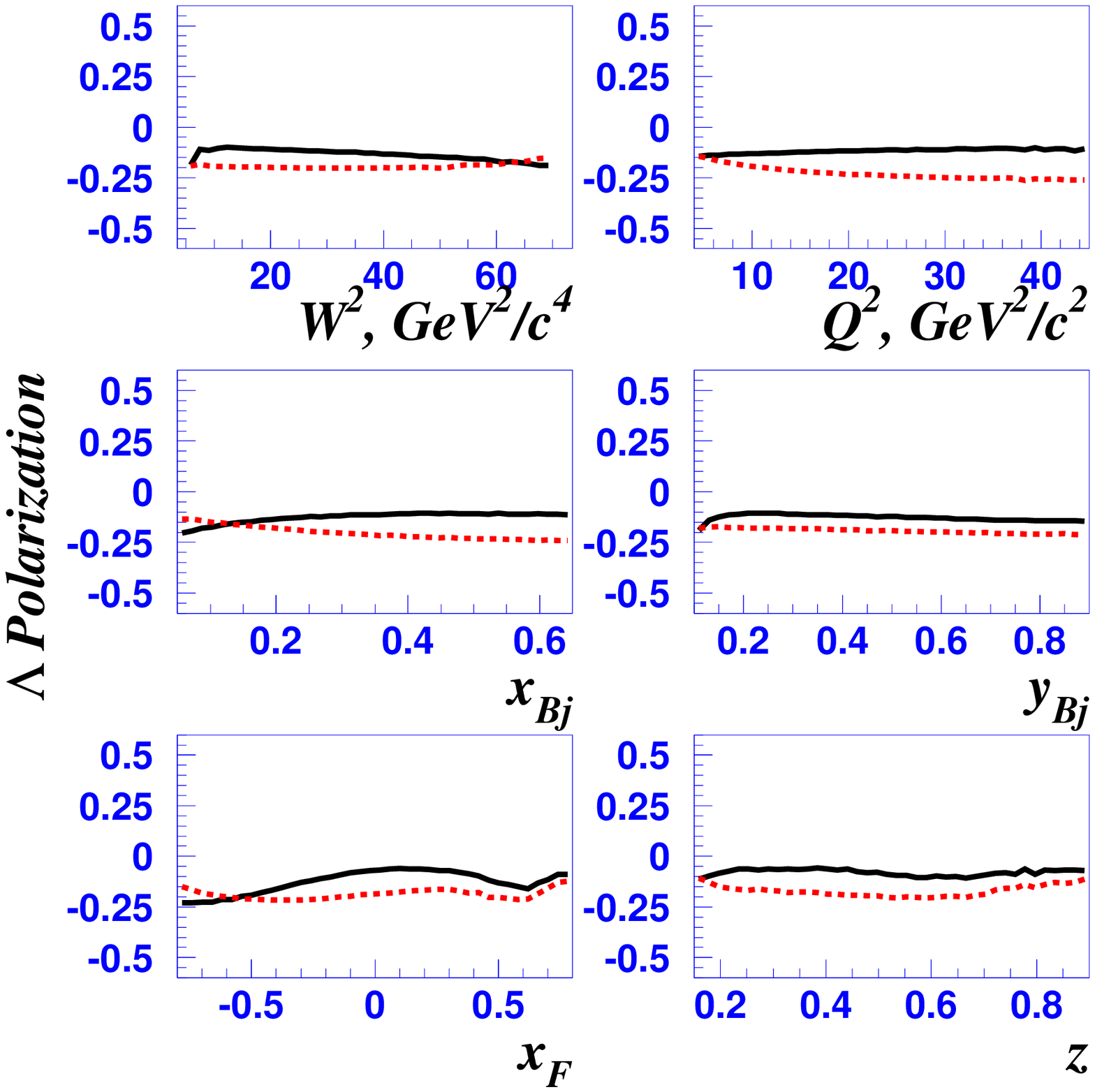,width=\linewidth}
\caption{\label{fig:numuxnc_6} The predictions of model A - solid line, 
model B - dashed line, 
for the polarization
of $\Lambda$ hyperons produced in $\nu_\mu$ neutral-current DIS
interactions off nuclei as functions of $W^2$, $Q^2$, $x_{Bj}$,
$y_{Bj}$, $x_F$ and $z$ (at $x_F>0$).} 
\end{center}
\end{figure}

\begin{figure}[htb]
\begin{center}
\epsfig{file=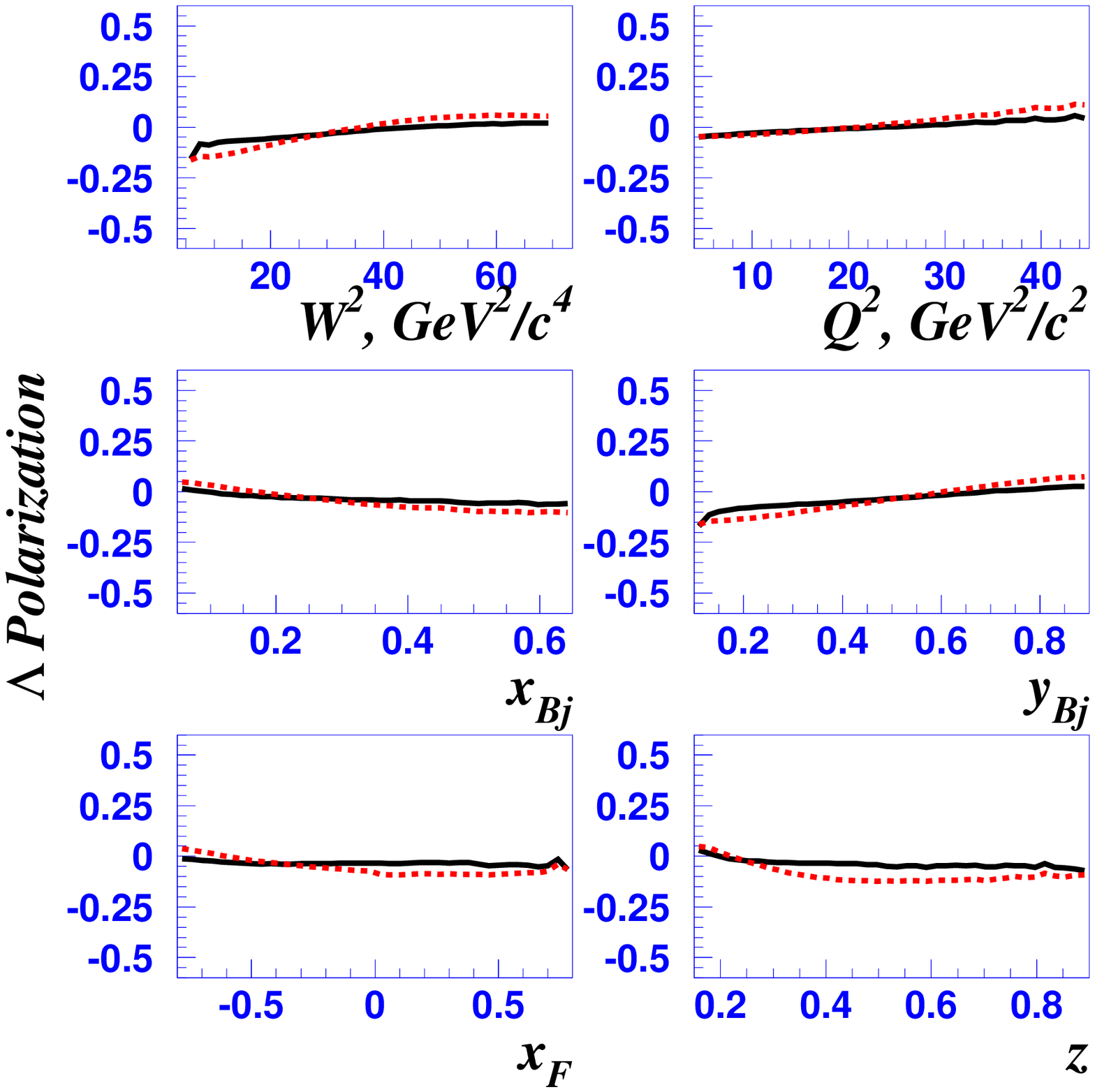,width=\linewidth}
\caption{\label{fig:anumunc_6} The predictions of model A - solid line, 
model B - dashed line, for the polarization
of $\Lambda$ hyperons produced in $\bar\nu_\mu$ neutral-current DIS
interactions off nuclei as functions of $W^2$, $Q^2$, $x_{Bj}$, 
$y_{Bj}$, $x_F$ and $z$ (at $x_F>0$).}
\end{center}
\end{figure}

\begin{figure}[htb]
\begin{center}
\epsfig{file=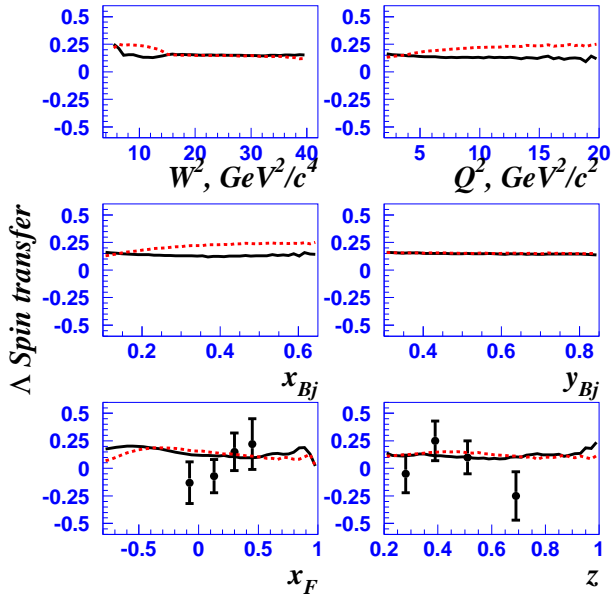,width=\linewidth}
\caption{\label{fig:e+xxxem_6} The predictions of model A - solid line, 
model B - dashed line, for the spin transfer to $\Lambda$ hyperons
produced in  $e^+$ DIS interactions off nuclei
as functions of $W^2$, $Q^2$, $x_{Bj}$, $y_{Bj}$, $x_F$ and $z$ (at
$x_F>0$). We assume $E_e = 27.5$~GeV, and the points with error bars are 
from~\cite{hermes}.} 
\end{center}
\end{figure}

\begin{figure}[htb]
\begin{center}
\epsfig{file=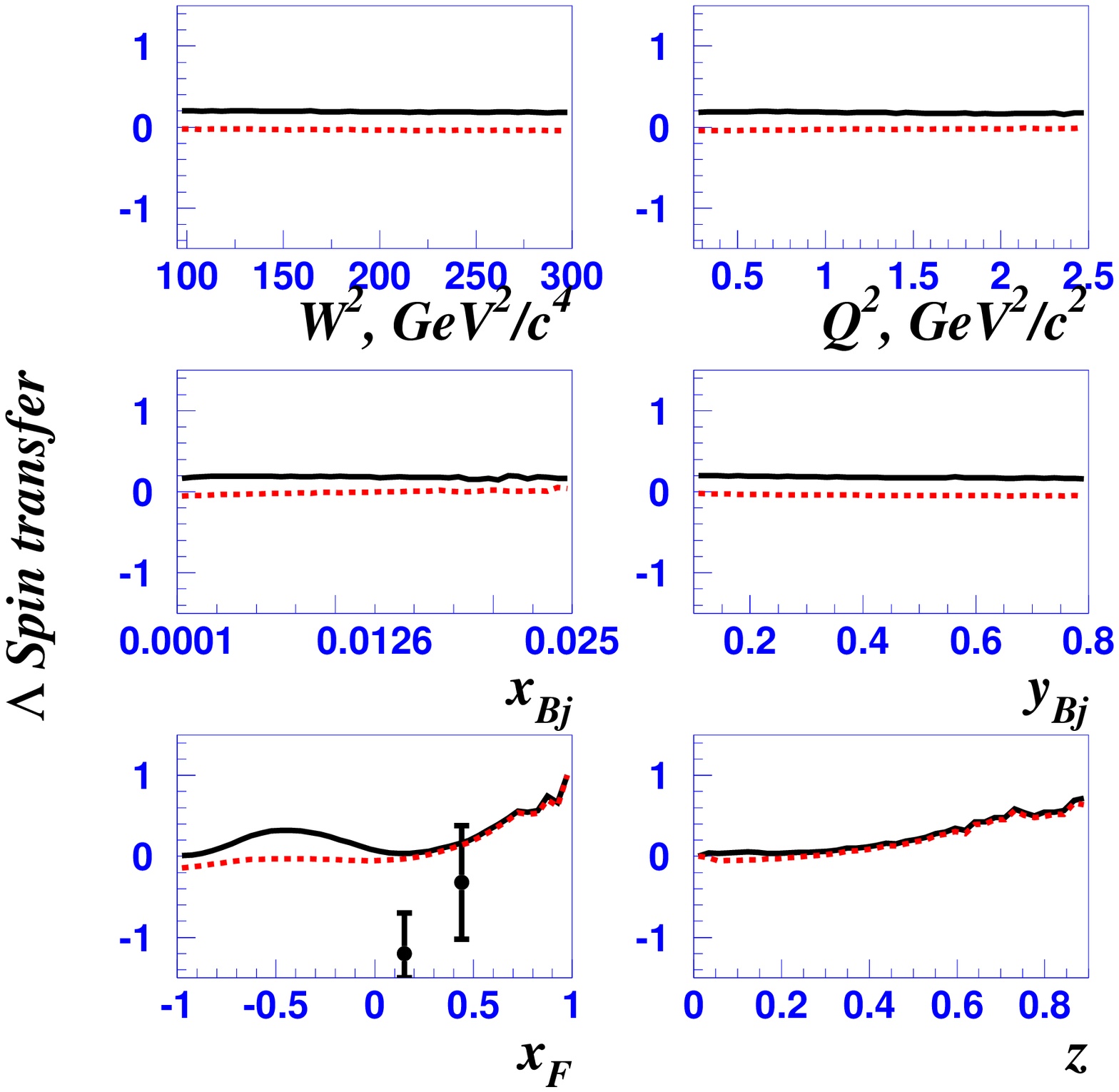,width=\linewidth}
\caption{\label{fig:e665_mu+xxem_6} The predictions of model A - solid 
line, model B - dashed line, for the spin transfer 
to $\Lambda$ hyperons produced in $\mu^+$ DIS
interactions off nuclei as functions of $W^2$, $Q^2$, $x_{Bj}$,
$y_{Bj}$, $x_F$ and $z$ (at $x_F>0$). Here we assume $E_\mu = 470$ GeV, 
as appropriate for E665.} 
\end{center}
\end{figure}

\begin{figure}[htb]
\begin{center}
\epsfig{file=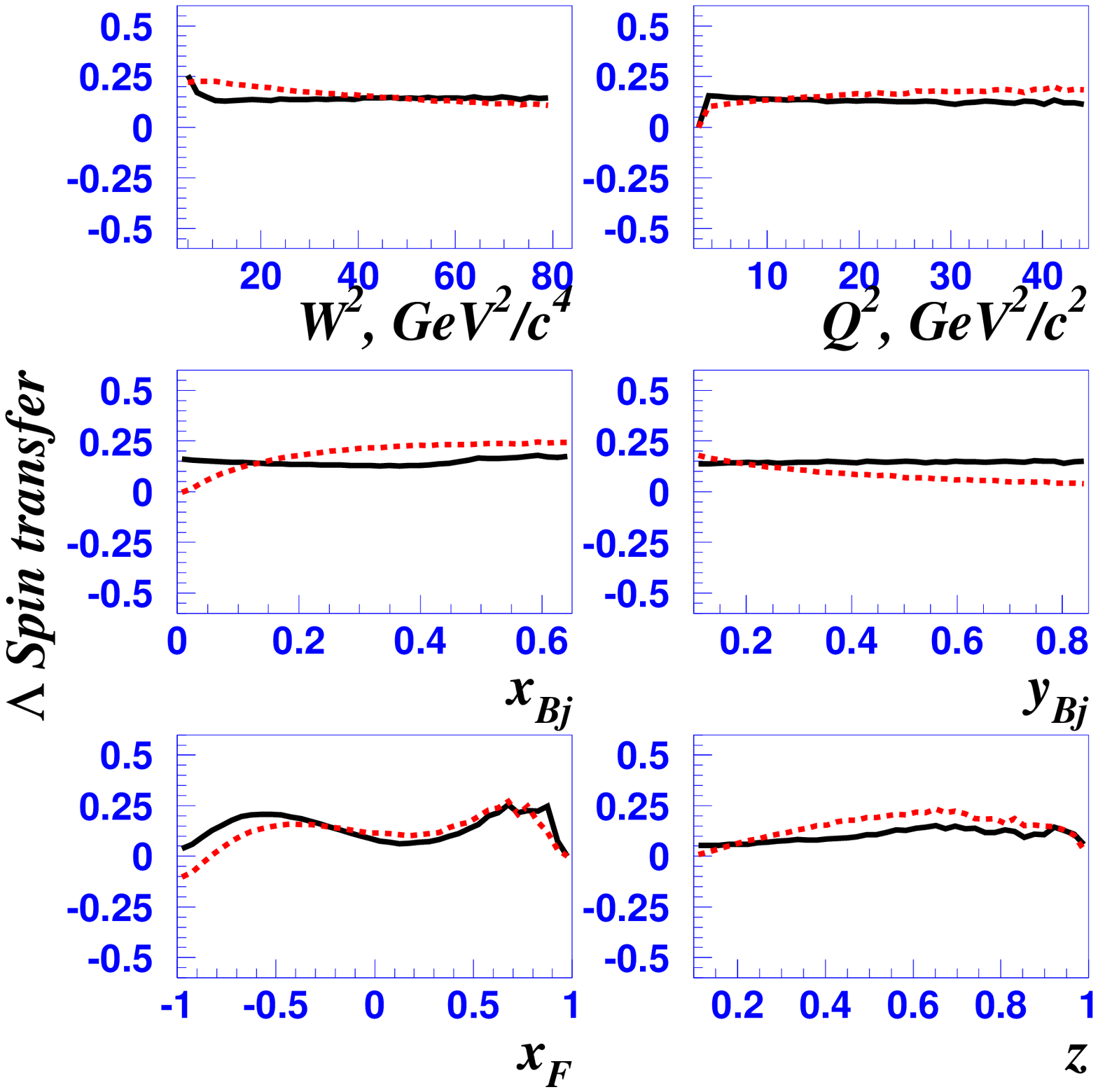,width=\linewidth}
\caption{\label{fig:compass_mu+xxem_6} The predictions of model A - solid 
line, model B - dashed line,
for the spin transfer to $\Lambda$ hyperons produced in $\mu^+$ DIS
interactions off nuclei as functions of $W^2$, $Q^2$, $x_{Bj}$,
$y_{Bj}$, $x_F$ and $z$ (at $x_F>0$). here we assume $E_\mu = 160$ GeV, 
as appropriate for COMPASS.} 
\end{center}
\end{figure}

%% file: all_tables.tex
\begin{table}[htb]
\begin{center}
\caption{\label{tab:nu-long-pol}\it  Longitudinal polarization of
  $\lam$ hyperons observed in bubble chamber (anti)neutrino experiments. }
\vspace*{0.5cm}
\begin{tabular}{c|c|c|c|c}
  \hline
  \hline
  Process          & $\langle E_\nu \rangle$ & & &\\
  Exp.(Year) & [GeV]    & cut on $x_F$ & $N_{\lam}$ & $P_x$\\
  \hline
  \hline
  $\nu_\mu - p$ & 51 &Full sample  & 289 & $-0.10 \pm 0.14$\\
  WA21~\cite{wa21}(1985) & & $x_F < 0 $ & 203 & $-0.29 \pm 0.18$\\
  & & $x_F >0 $ & 86 & $0.53 \pm 0.30$ \\
  $\bar \nu_\mu - p$ & 40 &Full sample  & 267 & $-0.24 \pm 0.17$\\
  WA21~\cite{wa21}(1985) & & $x_F <0 $ & 210 & $-0.38 \pm 0.18$\\
  & & $x_F >0 $ & 57 & $0.32 \pm 0.35$ \\
  $\bar \nu_\mu - Ne$ & 40 &Full sample  & 469 & $-0.56 \pm 0.13$\\
  WA59~\cite{wa59}(1992) & & $x_F <0 $ & 403 & $-0.63 \pm 0.13$\\
  & & $x_F >0 $ & 66 & $-0.11 \pm 0.45$ \\
  $\nu_\mu- Ne$(1994) & 150 &Full sample  & 258 & $-0.38 \pm 0.16$\\
  E632~\cite{e632} & & $x_F <0 $ & 190 & $-0.43 \pm 0.20$\\
  \hline
  \hline
\end{tabular}
\end{center}
\end{table}
~\\
~\\
\begin{table}[htb]
  \begin{center}
    \caption{\label{tab:pq} \it Polarization of the struck quark in 
different processes.}
    \begin{tabular}{c|c|c|c|c|c}
      \hline\hline
      & \multicolumn{2}{c|}{$\nu$} &  \multicolumn{2}{c|}{$\bar\nu$} &
      \multicolumn{1}{c}{$l^\pm$}\\   
      \cline{2-6}
      & CC & NC & CC & NC & $\gamma$  \\
      \hline
      &&&&&\\
      $P_d$ & $1$       & $\frac{(1-\xi)^2 - \xi^2(1-y)^2}{(1-\xi)^2 + \xi^2(1-y)^2}$
      & $-$
      & $\frac{(1-\xi)^2(1-y)^2 - \xi^2}{(1-\xi)^2(1-y)^2 + \xi^2}$
      & $P_B D(y)$\\
      &&&&&\\
      \hline
      &&&&&\\
      $P_u$ & $-$      & $\frac{(1-2\xi)^2 - 4\xi^2(1-y)^2}{(1-2\xi)^2 + 4\xi^2(1-y)^2}$
      & $1$
      & $\frac{(1-2\xi)^2(1-y)^2 - 4\xi^2}{(1-2\xi)^2(1-y)^2 + 4\xi^2}$
      & $P_B D(y)$\\
      &&&&&\\
      \hline\hline
    \end{tabular}
  \end{center}
\end{table}

\begin{table}[htb]
  \begin{center}
    \caption{\label{tab:cfr_coeff} \it Spin correlation coefficients in 
the $SU(6)$ and BJ models}
    \begin{tabular}{c|c|c|c|c|c|c}
      \hline\hline
      ${\lam}$'s parent & \multicolumn{2}{c|}{$C_u^{\lam}$} 
      & \multicolumn{2}{c|}{$C_d^{\lam}$} 
      & \multicolumn{2}{c}{$C_s^{\lam}$} \\  
      \hline\hline
      & SU(6) & BJ & SU(6) & BJ& SU(6) & BJ\\
      \cline{2-7}
      quark              & 0     & -0.18 & 0 & -0.18 & 1 & 0.63\\
      $\Sigma^0$         & -2/9  & -0.12 & -2/9 & -0.12 & 1/9 & 0.15\\
      
      $\Xi^0$            & -0.15  &  0.07 &  0   & 0.05 & 0.6 & -0.37\\
      $\Xi^-$            &  0    &  0.05 & -0.15 & 0.07 & 0.6 & -0.37\\
      ${\Sigma^\star}$   & 5/9   & -- & 5/9   & -- & 5/9   & -- \\
      \hline\hline
    \end{tabular}
  \end{center}
\end{table}


\begin{table}[htb]
  \begin{center}
    \caption{\label{tab:numuxcc}\it Dependence of the polarization of
      $\Lambda$  hyperons produced in $\nu_\mu$ CC DIS on the type of
target nucleon, compared 
      with the NOMAD~\cite{NOMAD_lambda_polar} data.}
    \input{numuxcc_NOMA.tex}
  \end{center}
\end{table}

%
\begin{table}[htb]
  \begin{center}
    \caption{\label{tab:anumucc}\it Model predictions for the polarization of
      $\Lambda$  hyperons produced in $\bar\nu_\mu$ CC DIS.}
    \input{anumucc_NOMA.tex}
  \end{center}
\end{table}
%
\begin{table}[htb]
  \begin{center}
    \caption{\label{tab:numuxnc}\it Model predictions for the polarization of
      $\Lambda$  hyperons produced in $\nu_\mu$ NC DIS.}
    \input{numuxnc_NOMA.tex}
  \end{center}
\end{table}
%
\begin{table}[htb]
  \begin{center}
    \caption{\label{tab:anumunc}\it Model predictions for the polarization 
of $\Lambda$  hyperons produced in $\bar\nu_\mu$ NC DIS.}
    \input{anumunc_NOMA.tex}
  \end{center}
\end{table}
%
\begin{table}[htb]
  \begin{center}
    \caption{\label{tab:e+xxxem}\it Model predictions for the polarization 
of $\Lambda$  hyperons produced in $e^+$ DIS with the HERMES  
experiment.}
    \input{e+xxxem_HERM.tex}
  \end{center}
\end{table}
%
\begin{table}[htb]
  \begin{center}
    \caption{\label{tab:e665}\it Model predictions for the polarization  
of $\Lambda$ hyperons produced in $\mu^+$  DIS with the E665 experiment.}
    \input{mu+xxem_E665.tex}
  \end{center}
\end{table}
%
\begin{table}[htb]
  \begin{center}
    \caption{\label{tab:compass}\it Model predictions for the 
polarization of $\Lambda$ hyperons produced in $\mu^+$  DIS 
with the COMPASS experiment, for $Q^2>1$ GeV$^2$, $x_F>-0.2$
      and $0.5<y<0.9$.}
    \input{mu+xxem_COMP.tex}
  \end{center}
\end{table}

%% file: numuxcc_NOMA.tex
\begin{tabular}{c|c|c|c}
\hline\hline
&\multicolumn{3}{c}{Target nucleon}\\
\cline{2-4}
$P_\Lambda$ (\%) & isoscalar & proton & neutron\\
\cline{2-4}
model A& -17.4& -11.4& -20.2\\
model B& -19.3& -18.1& -19.9\\
NOMAD& -15.0$\pm 3$ & -26.0$\pm 5$ &  -9.0$\pm 4$ \\
\hline\hline
\end{tabular}

%% file: anumucc_NOMA.tex
\begin{tabular}{c|c|c|c}
\hline\hline
&\multicolumn{3}{c}{Target nucleon}\\
\cline{2-4}
$P_\Lambda$ (\%) & isoscalar & proton & neutron\\
\cline{2-4}
model A&  -7.9& -10.7&  -3.4\\
model B&  -5.0&  -6.7&  -2.3\\
\hline\hline
\end{tabular}

%% file: numuxnc_NOMA.tex
\begin{tabular}{c|c|c|c}
\hline\hline
&\multicolumn{3}{c}{Target nucleon}\\
\cline{2-4}
$P_\Lambda$ (\%) & isoscalar & proton & neutron\\
\cline{2-4}
model A& -12.9& -11.7& -13.9\\
model B& -19.6& -19.6& -19.7\\
\hline\hline
\end{tabular}

%% file: anumunc_NOMA.tex
\begin{tabular}{c|c|c|c}
\hline\hline
&\multicolumn{3}{c}{Target nucleon}\\
\cline{2-4}
$P_\Lambda$ (\%) & isoscalar & proton & neutron\\
\cline{2-4}
model A&  -3.0&  -1.8&  -4.3\\
model B&  -3.1&  -2.8&  -3.4\\
\hline\hline
\end{tabular}

%% file: e+xxxem_HERM.tex
\begin{tabular}{c|c|c|c}
\hline\hline
&\multicolumn{3}{c}{Target nucleon}\\
\cline{2-4}
$P_\Lambda$ (\%) & isoscalar & proton & neutron\\
\cline{2-4}
model A&  -4.1&  -4.6&  -3.5\\
model B&  -4.3&  -4.4&  -4.1\\
\hline\hline
\end{tabular}

%% file: mu+xxem_E665.tex
\begin{tabular}{c|c|c|c}
\hline\hline
&\multicolumn{3}{c}{Target nucleon}\\
\cline{2-4}
$P_\Lambda$ (\%) & isoscalar & proton & neutron\\
\cline{2-4}
model A&  -4.7&  -4.7&  -4.7\\
model B&   1.1&   1.0&   1.2\\
\hline\hline
\end{tabular}

%% file: mu+xxem_COMP.tex
\begin{tabular}{c|c|c|c}
\hline\hline
&\multicolumn{3}{c}{Target nucleon}\\
\cline{2-4}
$P_\Lambda$ (\%) & isoscalar & proton & neutron\\
\cline{2-4}
model A&  -6.3&  -6.1&  -6.6\\
model B&  -3.0&  -3.1&  -2.8\\
\hline\hline
\end{tabular}